# Capturing Inner Experience At Scale

## An AI Interviewer Co-Developed with the Founder of a Landmark Phenomenological Method

*Jona Carmon*[1,2,3]*, Clara Bersch*[2,4]*, Charles Fernyhough*[5]*, Russell T. Hurlburt*[6]*, Simone Kühn*[1,7,8,9]

[1] Center for Environmental Neuroscience, Max Planck Institute for Human Development, Berlin, Germany
[2] Max Planck School of Cognition, Leipzig, Germany
[3] Department of Psychology, Humboldt-Universität zu Berlin, Germany
[4] Department of Industrial Automation Technology, TU Berlin, Berlin, Germany
[5] Department of Psychology, Durham University, Durham, UK
[6] Department of Psychology, University of Nevada, Las Vegas, USA
[7] Max Planck UCL Centre for Computational Psychiatry and Ageing Research, Berlin, Germany and London, UK
[8] Clinic and Policlinic for Psychiatry and Psychotherapy, University Medical Center Hamburg-Eppendorf, Hamburg, Germany
[9] Psychiatric Environmental Neuroscience (PEN) Clinic for Psychiatry CCM, Charité – Universitätsmedizin Berlin, corporate member of Freie Universität Berlin and Humboldt-Universität zu Berlin, Berlin, Germany

# Abstract

Subjective experience is central to psychological science, yet methods for studying it force a choice between depth and scale. Classical Experience sampling, as in ecological momentary assessments (EMA), captures experience as it occurs, but it confines participants to predetermined response formats that prescribe how experience is measured. Descriptive Experience Sampling (DES) instead investigates specific moments in depth through expert expositional interviews, but its reliance on scarce trained interviewers keeps samples small. Large language model (LLM) systems can scale qualitative interviewing. Some models operationalize established interviewing methods such as motivational interviewing, yet none is grounded in a method for apprehending inner experience. Here we present an AI interviewer that aspires to operationalize DES into an explicit, inspectable reasoning architecture. At each turn it appraises the participant's message across eleven quality dimensions, maintains a conservative account of what has been established, selects a stage-appropriate intervention, and composes a single non-leading query, always holding that temporal grounding precedes experiential content. It was derived from the full corpus of DES transcripts and refined with the method's originator Russell T. Hurlburt. To our knowledge it is the first AI interviewer grounded in an established method for studying inner experience. The interviewer runs inside *Introscope*, an application that delivers the beeps and conducts the interviews and a study platform that lets researchers run studies via shareable links and review the sampled experience. It is demonstrated in an accompanying video https://introscope.mpib-berlin.mpg.de/video. Pending validation studies, we will make it freely available to researchers and the public, for crowdsourced sampling and individual exploration of inner experience.

# Introduction

Inner experience, what a person is thinking, feeling, sensing, and so on, at a given moment, is central to psychological science, from theories of consciousness to clinical assessment. Experience sampling methods (ESM) and ecological momentary assessment (EMA) have become dominant approaches to capturing experience as it occurs, using prompts to sample participants at random moments throughout daily life (Csikszentmihalyi & Larson, 2014; Shiffman et al., 2008). But these methods require participants to report within the boundaries of predetermined categories, such as Likert scales, fixed emotion labels, and categorical checklists, that prescribe how experience is conceptualized and measured, and do not allow new, potentially idiosyncratic, forms of experience to be captured. Research on introspection has repeatedly demonstrated that people have limited access to mental processes underlying their behavior and that self-reports of those processes are shaped by prior theories and expectations rather than direct observation (Nisbett & Wilson, 1977). This problem is compounded when measurement instruments constrain reporting to predetermined categories, because respondents use response formats themselves as informational cues about what kind of answer is expected (Schwarz, 1999). In experience sampling, the result is that researchers risk recovering the structure they built into their instruments rather than the structure of experience itself.

Descriptive approaches to experience sampling address this problem by replacing fixed-response formats with open-ended, in-depth investigation of specific moments. Rather than selecting from predetermined categories, a skilled interviewer helps participants articulate what was present in their experience at a given moment, often revealing phenomena that no checklist would have anticipated and allowing for idiosyncratic experiences that are unique or at least not frequently observed between individuals. The method most fully developed for sampling inner experience as it naturally occurs, moment to moment, in everyday life, is Descriptive Experience Sampling (DES; Hurlburt & Akhter, 2006). It samples moments randomly identified by a beeper and directs the expositional interview at a single specified moment, the precise moment caught in flight by the beep, following explicit principles such as open-beginninged (as well as open-ended) first questions, non-leading follow-ups, bracketing of both investigator and participant presuppositions, and rigorous focus on that moment, with participants refining their descriptive skill across repeated sampling days. Using DES, frequently occurring (but non-exhaustive) phenomena of inner experience have been identified, including inner speaking, inner seeing, unsymbolized thinking, feelings, and sensory awareness, with substantial individual variation (Heavey & Hurlburt, 2008).

However, the depth of engagement that descriptive experience sampling affords, and its capacity to detect unexpected experiential forms, comes at the cost of scalability. DES requires trained expert interviewers and extensive one-on-one interaction, which is why sample sizes rarely exceed n = 10 per study.

The emergence of large language model-based systems has created a potential path beyond this constraint. Recent work has demonstrated that AI systems can conduct semi-structured qualitative interviews, uncovering multifaceted patterns that go beyond participants' initial top-of-mind responses; using LLM systems enables sample sizes large enough for quantitative subgroup comparison (Chopra & Haaland, 2023). This approach has been shown to generalize across diverse domains, including decision-making, political views, and subjective mental states, with interview quality rated comparable to an average human expert (Geiecke & Jaravel, 2024). A small number of systems go further and operationalize a specific established interviewing method. LLM-powered agents have implemented motivational interviewing for health counseling (Steenstra et al., 2024) and the cognitive interview for forensic witness recall (Minhas et al., 2022). These are important precedents, but their target is behaviour change and accurate memory retrieval, not the apprehension of inner experience. More generally, AI interviewers to date are created to implement general best practices drawn from sociological and survey methodology, such as conducting interviews non-directively, encouraging elaboration, and gathering concrete examples. They are not grounded in a method for studying subjective experience, and none was built in collaboration with the originator of the method it implements. The qualitative research community's concerns about unprincipled AI deployment (Jowsey et al., 2025), while primarily directed at AI-driven analysis, reinforce the point that methodological rigor is not something scaling should bypass.

Here, we address this gap by developing an AI interviewer grounded in DES methodology. DES interviewing requires ongoing, simultaneous assessment of multiple quality dimensions across an interview, from temporal grounding to bracketing of presuppositions to progressive clarification of experiential content. Hurlburt and colleagues have made several hundred hours of DES interviews, as both video and transcript, publicly available (Hurlburt, 2026). We initially took a corpus-driven approach, instructing a language model to read this corpus and interview in the style of its originator, but the resulting interviews with volunteer participants did not replicate expert practice: the model strayed from the precise moment of the beep, failed to detect unclear descriptions, was too easily satisfied with the answers it received, and tried to please the participant where DES calls for the opposite.

We therefore shifted to a framework-driven design. Rather than having the model imitate the corpus, the authors, together with the method's originator, specified an explicit reasoning framework: a fixed order of rationally derived, scorable judgments the interviewer must perform before it speaks. The model executes this framework, using the corpus as evidence rather than as the reasoning engine. The stages and scoring criteria were fixed by five decades of DES experience, while the judgments within them are made by the model, so the interviewer is a hybrid of human-rational design and AI computation. We embed it in *Introscope*, an application and study platform that lets DES run at scale in participants' daily lives. To our knowledge, this is the first AI interviewer grounded in an established psychological method for the study of inner experience, and the first co-developed and reviewed by that method's originator. We do not claim it is a complete, final, or fully adequate operationalization of DES; formal validation against expert-conducted interviews is the next step. In this paper, as a preliminary but important step in the creation of such methods, we present the architecture, the operationalization of expert DES practice into explicit and scorable criteria, and a qualitative formative evaluation of the AI interviewer's fidelity to expert practice.

# Results

## System Architecture

A DES expositional interview is not a script but a dynamic, assessment-driven process. At every turn the interviewer (whether human or AI) must judge how well the participant has isolated the precise instant of the beep, judge how well the participant constrains themself to describing experience and nothing else, judge the quality of the experiential description itself, and maintain a running and deliberately conservative account of what has actually been established about the beeped experience. Out of all that, the interviewer composes the next query, appropriate to the interview's stage, aimed to be a non-leading question in the participant's own language.

For an AI interviewer, the running account is formalized as the experience registry, described in Stage 2 below. The AI interviewer described here renders each of these judgments, from isolating the moment of the beep to composing the next query, as explicit, overlapping scorable steps, with one methodological commitment encoded throughout. For example, temporal grounding is a prerequisite for experiential content – until the beeped moment is located, any account of what was there is treated as provisional; and until experience is grasped, any account of when an experience might have occurred is treated as tentative.

To incorporate those features, we developed a reasoning architecture that comprises four stages: Message Appraisal, Experience Modeling, Strategy Selection, and Query Composition (Figure 1). Each participant-message triggers one call to the LLM, which executes the four stages as ordered steps within its extended thinking, streams the plain-text query the participant sees, and then emits a single structured analysis object (via a tool call) carrying the appraisal, the updated experience registry, the selected intervention priorities, and the session's closure recommendation. The four reasoning stages therefore run inside that single call, directed by one system prompt (the fixed set of instructions that governs the model's behaviour, as distinct from the participant's message) whose numbered sections instruct the model to perform the stages in order before producing any output.

**Stage 1. Message Appraisal.** Each message is categorized as core (an experiential description), phatic (small talk, hedging, acknowledgment), or meta (a question about the method, logistics, or an ethics boundary). Only core messages are scored. Core messages are scored on eleven quality dimensions (split into two groups), 3 moment-of-the-beep dimensions and 8 experiential-quality dimensions.

| Dimension | Group | What it assesses |
|---|---|---|
| Leading-edge clarity | Moment | Whether the description is anchored to the precise moment of the beep or spread across a broader time span |
| Before/after separation | Moment | Whether the participant distinguishes what was present at the beep from retrospective reflection that followed it |
| Sequence clarity | Moment | Whether the temporal ordering of multiple reported elements is discernible |
| Modality clarity | Experiential | How precisely the experiential form is identified as for example inner speaking, inner seeing, feeling, sensory awareness, or unsymbolized thinking, if those are present (Heavey & Hurlburt, 2008) |
| Interpretive unpacking | Experiential | Whether vague everyday terms such as "thinking" or "realizing" have been clarified into specific experiential content |
| Concrete detail | Experiential | Presence of specific experiential particulars such as exact words, visual properties, or bodily locations |
| Source clarity | Experiential | Whether what was directly apprehended is distinguished from what is merely known or inferred |
| Coherence | Experiential | Internal consistency across described elements |
| Metaphor specificity | Experiential | Whether figurative language has been unpacked into literal experiential content |
| Salience | Experiential | Whether the participant has identified what was central versus peripheral |
| Separability | Experiential | Whether multiple reported aspects are one integrated experience or distinct co-present phenomena |

Each dimension has to be turned into a quality score. Rather than having the model produce that number directly, it selects one of a small set of named, ordered categories, each defined by a short description and mapped to a fixed place on the scale, much like the levels of a grading rubric. The categories run from one capturing the ideal for that dimension down to one capturing its worst case, with a separate label for messages that have nothing to judge on that dimension. Choosing a category rather than a number keeps scoring consistent across messages and keeps each score traceable, because the category names the quality being judged rather than leaving an unexplained figure. These scores decide where the interviewer intervenes next. An intervention is a strategic move that targets one dimension in order to sharpen it, that is, to get the participant to clarify that aspect; the interviewer realizes each intervention as its next query. Dimensions that fall short are flagged as candidates, but the interviewer intervenes only on the genuinely weak ones, so it addresses what is actually unclear rather than perfecting every dimension at once.

At each turn the model also makes one judgment that is never scored or targeted by an intervention: the participant's experiential access, meaning how directly they are in contact with the experience itself rather than merely talking about it. We call this a mediating dimension because it conditions how the interviewer handles all of the quality dimensions at once, and how hard it probes. It follows one explicit rule, namely that fluency is not access: a report or generic mental label such as "I was thinking about my brother," however confidently stated, counts as low access until concrete phenomenal detail emerges and the moment is at least roughly located. The model must justify each access judgment in its own reasoning, pointing to the specific hedging markers and temporal cues behind it, so the judgment stays grounded and inspectable rather than arbitrary. Separately, the model tracks three further signals that are not mediating dimensions but still shape the interview: the participant's fatigue, the emotional sensitivity of the content, and a closure signal indicating that the moment has been explored enough to end.

**Stage 2. Experience Modeling.** This stage maintains the experience registry, which is a progressive structure with three buckets: directly-in-experience (= confirmed at the beep), uncertain (= not yet resolved), and not-in-experience (= ruled out). The experience registry is designed to be conservative: new information defaults to uncertainty and is promoted to directly-in-experience only when clearly established through the interview. This conservatism is itself modulated by access judgment. When access is low, only coarse, provisional entries are permitted, so the interviewer does not encode complex experiential structure as established fact while the participant's own reports remain tentative. The registry is carried back in on the following turn, so understanding is built incrementally rather than re-derived from scratch, mirroring how a skilled interviewer accretes understanding rather than accepting initial reports at face value.

**Stage 3. Strategy Selection.** The AI interviewer selects one to three intervention priorities from the appraisal scores and the current registry state. That selection depends on the current interview stage.

Early in the sampling sequence, moment dimensions are prioritized regardless of experiential form scores, because securing the moment is prerequisite to any meaningful exploration of content. If participants reliably anchor to the moment, subtler dimensions such as source clarity and modality take precedence. An anti-circularity mechanism tracks intervention history. If the same intervention has been tried repeatedly without score improvement, the AI interviewer pivots to a different dimension or escalates with a different approach, and several consecutive attempts on one dimension force a pivot or a recognition of completion. Once priorities are set, the AI interviewer draws on a strategy repertoire that supplies concrete guidance for each strategy type, adapted to the participant's current access level.

**Stage 4. Query Composition.** The query is generated under the usual-to-DES constraints. It uses non-leading phrasing that presupposes nothing about what kind of experience was present (which DES calls "open-beginninged"), for example, asking what if anything was there rather than what the participant was feeling. It stays in the participant's own language rather than importing interviewer terminology, asks one question at a time to protect introspective focus, and hedges when access is low, signaling that the interviewer's understanding is tentative and inviting correction rather than confirmation. It explicitly prohibits confirming the participant's interpretations, offering generic praise (countering the documented sycophancy of large language models; Sharma et al., 2024), or overriding "I don't know", which in DES is treated as valuable epistemic information rather than a failure of recall. The query is plain text that the model writes and streams to the participant first, before it records the structured analysis which speeds up the time until the query is sent to the participant.

Two safeguards sit above the four stages. A wellbeing layer takes priority over every interviewing goal: whenever the participant shows signs of present distress, as distinct from merely reporting negative experience at the beep, the AI interviewer is instructed to pause the method, gently offer to end, and point the participant to the human researchers, whose contact details are always available in the app. A security layer treats everything the participant writes as data to be analyzed rather than instructions to be followed, so attempts to change the interviewer's role or draw out its internal state are to be declined and redirected without interrupting the interview.

**Between sessions, the Participant Profile.** After a session closes, a profile generator digests the interview into a longitudinal record covering the participant's access-level distribution, intervention history, and descriptive ability. At the next session, this profile is formatted back into the system prompt (the fixed instructions the AI interviewer works from), so the AI interviewer adapts to a returning participant and supports the iterative skill-building that DES intends. Because the system prompt is static and cached, only the evolving state is supplied per turn, namely the previous registry, recent intervention history, the prior classification state, turn number, elapsed time, the profile, and the new message.

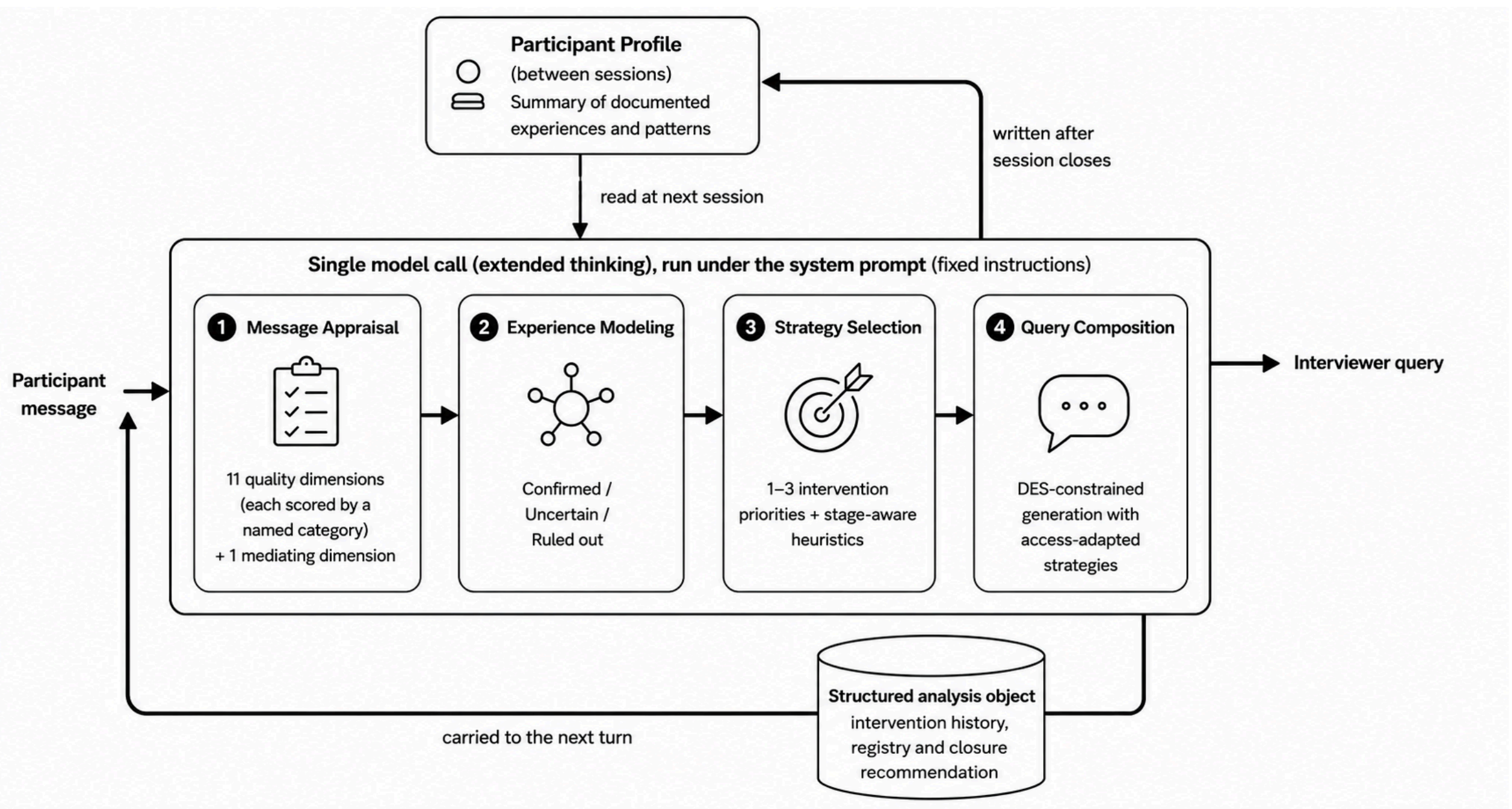

***Figure 1: Reasoning architecture of the DES AI interviewer.*** *Each participant message triggers a single model call that executes four logical stages in order, namely Message Appraisal, Experience Modeling, Strategy Selection, and Query Composition, within the model's extended thinking, before streaming a plain-text query and emitting a structured analysis object. The Participant Profile operates between sessions and is formatted back into the system prompt to inform subsequent interviews.*

## Iterative evaluation

The AI interviewer scheme that we have been discussing was created and developed through a structured iterative process comprising five phases (Figure 2). We treat the later phases not merely as development steps but as formative evaluation. Each provides evidence about the degree to which the interviewer reproduces expert DES practice, and about the failure modes that remain. Because no quantitative outcome benchmark for DES interview quality previously existed, and indeed the appraisal dimensions reported here are the first such operationalization, this evaluation is qualitative and convergence-based.

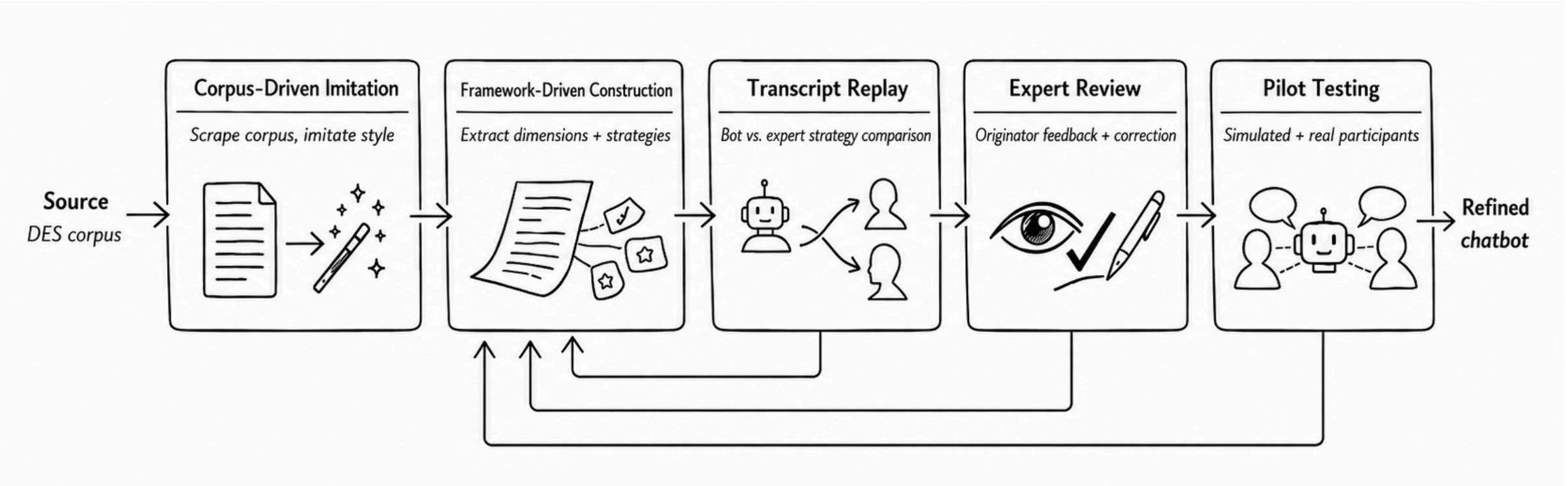

***Figure 2: Iterative development and formative-evaluation process.*** *The AI interviewer was iteratively refined across five phases, namely Corpus-Driven Imitation, Framework-Driven*

*Construction, Transcript Replay, Expert Review, and Pilot Testing. Development proceeded from the complete corpus of documented DES interview transcripts, with each phase producing revisions carried forward into subsequent phases.*

**Corpus-Driven Imitation.** As a starting point, a language model was instructed to draw on the documented DES corpus and interview in the style of the method's originator, with no explicit decomposition of the method. Tested with volunteer participants, this baseline did not satisfactorily replicate expert DES practice: among other shortcomings, it strayed from the moment of the beep, failed to notice unclear descriptions, was too readily satisfied with the answers it received, and tended to please the participant where DES calls for the opposite. These failures motivated the Framework-Driven Construction described next.

**Framework-Driven Construction.** The complete corpus of documented DES interview transcripts, provided by the method's originator, was analyzed to extract the quality dimensions and intervention strategies that constitute expert DES practice. The eleven appraisal dimensions and the intervention strategy repertoire were derived from this corpus. This operationalization phase is itself a contribution, rendering previously tacit expert judgment as an explicit, inspectable set of scorable criteria.

**Transcript Replay.** Participant turns from existing expert transcripts were fed to the AI interviewer, and its selected strategies were compared to those actually employed by the human interviewer at the corresponding point in the interview. Across iterations of revision, early divergences were eliminated. For example, we adjusted the model to require that  the system probed temporal aspects before experiential aspects.

**Expert Review.** The method's originator (R.T.H.) reviewed system-generated transcripts at multiple stages and provided feedback that clarified methodological intent and corrected misalignments. This review functioned as an expert-validity check on the operationalization itself, confirming that the appraisal dimensions and intervention strategies faithfully reflect the method as practiced.

**Pilot Testing.** The interviewer was tested with LLM-simulated participants and with researchers within and outside the project team acting as real participants. This stage identified issues not visible in transcript replay.

## The application

Around the AI interviewer sits Introscope, the application and study platform that lets a researcher run the AI-DES in participants' everyday lives without a human interviewer in the loop (Figure 3). A researcher signs in, writes a privacy notice and consent text, sets a data-retention period, and defines a beep plan covering how many days, how many beeps per day, the daily hour window, and whether sampling dates are fixed or chosen by the participant within a span. Participants are onboarded passwordlessly. The researcher shares an invite link, and the participant installs a native iOS app and claims the link, which issues a device token tied only to an anonymous participant code, with no name and no password. After giving consent, and after selecting their sampling days, the participant's device is scheduled with the requested number of beeps placed at random times within the daily window, in the participant's own timezone, for each sampling day. When a beep is due, it is delivered as a push notification with a distinguished DES beep, and participants can record their notes and complete the DES interview by typing or by speaking and listening, on either a phone or a computer. During the interview, each turn streams a query and saves its structured analysis until the AI interviewer detects closure. The researcher's dashboard then shows, for each participant, the beep outcomes (delivered, failed, upcoming), interview counts (completed, open), the full transcripts, the evolving experience registry (items confirmed, uncertain, or excluded), and the current participant profile. Underneath, every database query is scoped by row-level security to the current user, researcher and participant authentication are kept separate, and a governance layer enforcing consent, and retention wraps the whole system. This architecture is what turns the AI interviewer from a single conversational agent into an instrument for population-scale experience sampling. Many participants can be sampled and interviewed concurrently, immediately after each beep, with no expert interviewer time consumed per interview.

Introscope is built for researchers running their own studies, each under their own API keys (see Methods). The same architecture also enables public, crowdsourced sampling of inner experience. Following the validation studies, we intend to host an open deployment of Introscope under our own researcher keys.

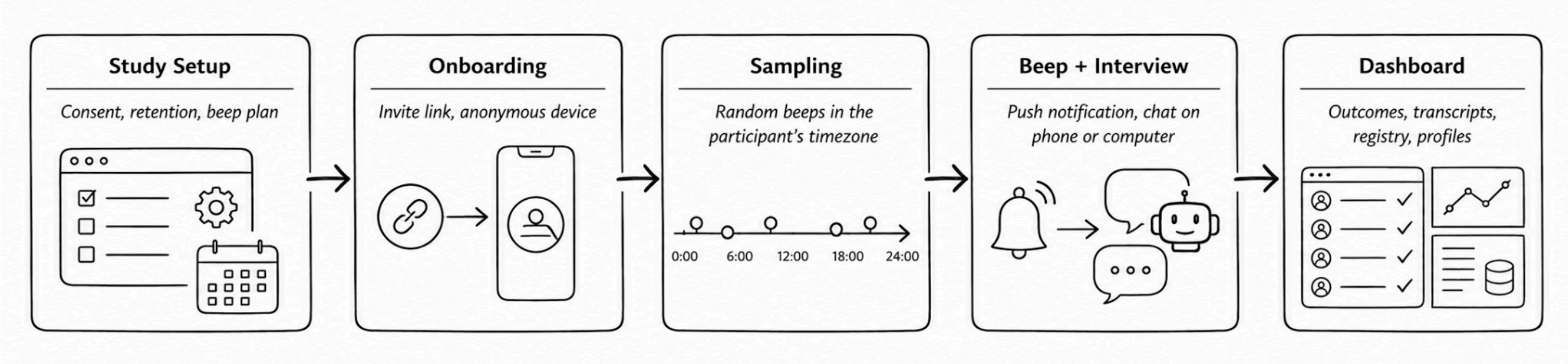

***Figure 3. Introscope, the study platform and participant app.*** *A researcher configures a study and reviews results on a dashboard while participants are sampled and interviewed in their daily lives, with a governance layer of consent, retention, and row-level isolation underlying every stage.*

# Discussion

We have developed and qualitatively evaluated what is, to our knowledge, the first AI interviewer grounded in an established psychological method for the study of inner experience. The AI interviewer reproduces, in an explicit and inspectable form, interviewing behaviour that until now has resided only in the trained judgment of a small number of expert practitioners.

This contribution builds on a fast-moving literature. Recent work has established that LLM-based systems can conduct qualitative interviews at scale with data quality rated comparable to human interviewers (Chopra & Haaland, 2023; Geiecke & Jaravel, 2024), and a handful of systems operationalize specific established methods such as motivational interviewing (Steenstra et al., 2024) and the cognitive interview (Minhas et al., 2022). We extend this trajectory into inner experience, where the interviewing demands, namely temporal grounding to a single instant, bracketing of presuppositions, and the treatment of uncertainty as information rather than failure, differ substantially from those of behaviour-change or memory-retrieval interviewing, and where generic interviewing heuristics are accordingly least adequate.

The architecture also gives something back to the study of inner experience itself, independent of the application. The eleven quality dimensions aspire to operationalize what was previously implicit expert knowledge about DES interview quality, making that knowledge explicit and inspectable, and the experience registry aspires to formalize the progressive, conservative epistemic posture that skilled interviewers maintain. These operationalizations may prove valuable on their own, for instance as training tools for investigators of inner experience or as evaluation frameworks for interview-quality research.

A further distinctive feature is that the AI interviewer was built and reviewed in direct collaboration with the originator of the method it implements. This both increases the likelihood that the operationalization reflects the method as intended and sets the AI interviewer apart from prior method-grounded interviewers, which drew on the published literature rather than the originator's own corpus and review. Grounding the interviewer in the complete documented transcript corpus, rather than in secondary descriptions, was essential to capturing practices that are easier to demonstrate than to state.

That involvement also speaks to a wider unease. The qualitative research community has raised concerns about unprincipled AI deployment in research (Jowsey et al., 2025). Our work answers the spirit of these concerns. Rather than deploying AI generically, we constrain it within an established method, make its decision criteria inspectable, and involve the method's originator throughout development. As described in the Results, the AI interviewer is a carefully human-designed and evaluated structure rather than a naive application of AI to interviewing. Constraining it in this way also means working against the model's own defaults. A generic conversational model tends toward sycophancy, agreeing with and validating the user rather than questioning them (Sharma et al., 2024), whereas DES demands the opposite, so much of the design effort went into making the AI interviewer question and withhold agreement where an unconstrained model would confirm. Methodological grounding of this kind may turn out to be the kind of thing that separates responsible AI integration from generic deployment.

Beyond scale, the architecture makes possible a structural change in how DES interviews are conducted. Traditional practice requires participants to collect about six beeps over a day and describe them all in a single interview the following day, a structure dictated by the logistics of scheduling expert interviewer time and by the difficulty of remembering more than six beeps, not by methodological preference. DES has always acknowledged that the delay between beep and interview opens room for memory distortion and post-hoc interpretation (Hurlburt & Akhter, 2006), a concern reinforced by broader evidence that retrospective reports of experience diverge systematically from momentary assessments even over short intervals (Shiffman et al., 2008; Stone & Shiffman, 2002) but reduced by the DES reliance on iterative training. Because the AI interviewer delivers an interview immediately at the beep, it removes this constraint in principle, allowing participants to be interviewed while the experience is maximally fresh. Whether that temporal proximity improves the fidelity of experience reports is an empirical question the AI interviewer now makes testable.

## Limitations

This paper reports the interviewer and a qualitative formative evaluation, but it does not establish that the AI interviewer's interviews match those of experts at the level of experiential outcomes, which warrants further investigation in upcoming validation studies with phenomenological experts of different traditions. Two further limitations bear on the present contribution. First, the operationalization was reviewed by R.T.H., the originator and most experienced practitioner of the method, which anchors the system to best-available practice but reflects one practitioner's interviewing style. Second, the AI interviewer accepts only verbal input. Interviews are conducted in text or, when the researcher enables it, in speech, but in both cases the participant must put experience into words. DES does not in principle constrain response modality: it allows figures or diagrams alongside speech, even if in practice responses are overwhelmingly verbal. Confining input to words may favour more verbally articulate participants and disadvantage those whose experience is less readily put into words, a consideration that may bear particularly on some neurodivergent participants. Since nothing in principle prevents an AI interviewer from handling figures or diagrams, accepting them is a natural extension for future versions.

### Future directions

A method-grounded, scalable interviewer opens several avenues. DES has a long history of application in clinical populations, including individuals with schizophrenia, depression, anxiety disorders, and bulimia, revealing clinically relevant features of inner experience that standardized instruments miss (Hurlburt, 1993, 2012), as well as to neurodivergent populations such as adults with Asperger syndrome (Hurlburt et al., 1994). A validated interviewer could enable more frequent experience monitoring than expert-led methods can sustain. It could also support the study of experience in context, that is how physical and social environments shape inner experience (Kaplan, 1995), and, combined with physiological and neuroimaging measures, the study of experience and body relationships at a scale that has so far been limited to small case studies (Fernyhough et al., 2018; Hurlburt et al., 2015). Realizing any of these directions rests first on formally validating the system's interviews against expert-conducted DES interviews, the immediate next step for this work.

### Conclusion

By grounding AI interviewing in an established psychological method and rendering its decision criteria explicit and inspectable, we demonstrate that the systematic, in-depth study of inner experience need not remain confined to the small samples that expert-led methods permit. The AI interviewer does not aim to replace the expert human interviewer but to extend the reach of the method and to make its tacit expertise public. This work offers a path between uncritical adoption of AI tools and blanket rejection of their potential to advance the study of human experience.

## Methods

### Descriptive Experience Sampling.

DES is a method for exploring inner experience developed by Hurlburt (Hurlburt & Akhter, 2006; Hurlburt & Heavey, 2006). Participants carry a random beeper in their daily life. When the beep sounds, they attend to whatever was in their experience at that moment and jot down brief notes. For logistic reasons, participants typically collect six beeps over the course of a day and describe all of them in a single expositional interview with a trained investigator on that same or the following day. The interview follows principles that distinguish it from standard qualitative interviewing, including “open-beginninged” first questions that avoid presuppositions about what (if any) experiences might have been present, non-leading follow-ups, bracketing of both the investigator's and the participant's assumptions about experience, and rigorous focus on the specific moment of the beep rather than generalizations about what the participant typically experiences. Across multiple sampling sessions, both investigator and participant iteratively improve their skill at apprehending and describing experience (Hurlburt & Akhter, 2006).

### Technical implementation.

The reasoning architecture and the study platform Introscope are described in the Results. The AI interviewer runs on Anthropic's Claude, with interviews and between-session profile generation handled by Claude Sonnet 4.6. The architecture is model-agnostic and has also been run on other sufficiently capable models during development. Each researcher supplies their own API key and chooses the provider, either the Anthropic API directly or an EU-region Azure AI Foundry deployment for EU data residency, so participant data is processed under the researcher's own data-protection arrangements. Voice interviewing (speech-to-text input and a spoken-conversation mode with text-to-speech, powered by Mistral Voxtral) is available as a per-study option when the researcher supplies a voice key; otherwise interviews are text-only.

## Acknowledgements

The authors used Claude Opus 4.8 (Anthropic) and GPT 5.6 (OpenAI) as a writing assistant during manuscript revision and DALL-E 3 (OpenAI) to generate figure graphics. The application was built with the support of Claude Code (Anthropic) as a coding assistant, under human planning, testing, revision, and safety review. All intellectual content was developed, written, and approved by the authors.